\documentclass[english,keywords,showpacs]{iopart}
\usepackage[pdftex]{graphicx}
\usepackage{mathrsfs}
\usepackage{epstopdf}
\usepackage{epsfig}
\usepackage{amssymb}
\usepackage{xcolor}
\usepackage{iopams}
\usepackage{upgreek}
\usepackage{bbold}
\usepackage{bm}
\usepackage[colorlinks, citecolor = blue, urlcolor = blue]{hyperref}
\usepackage[square,sort&compress,comma,numbers]{natbib}
\usepackage{cite}
\usepackage{longtable}
\usepackage{har2nat}  % loads 'natbib' automatically
\usepackage[titletoc,title]{appendix}
\usepackage{caption}
\usepackage{braket}
\usepackage{xcolor}
\usepackage{todonotes}
\begin{document}

\title{Quantum conditional mutual information of W state in non-inertial frames}

\author{H Saveetha$^1$ \footnote{saveehari@gmail.com},
Peter P. Rohde$^2$,$^3$ \footnote{dr.rohde@gmail.com} and
R Chandrashekar$^4$,$^5$ \footnote{cr2442@nyu.edu}}
\address{$^1$ Centre for Quantum Science and Technology, 
\\Chennai Institute of Technology, Chennai 600069, India}
%{https://www.peterrohde.org}
\address{$^2$ Centre for Quantum Software \& Information (UTS:QSI), \\University of Technology Sydney, Ultimo, NSW 2007, Australia}
\address{$^3$ Hearne Institute for Theoretical Physics, Department of Physics \& Astronomy, Louisiana State University, Baton Rouge LA, United States}
\address{$^4$Department of Computer Science and Engineering, NYU Shanghai, 567 West Yangsi Road, Pudong, Shanghai 200124, China}
\address{$^5$ Centre for Quantum Information, Communication and Computing, \\Indian Institute of Technology Madras, Chennai 600036, India}

\vspace{10pt}

\begin{abstract}
Quantum conditional mutual information (QCMI) is a versatile information theoretic measure.  It is used to find the amount of correlations 
between two qubits from the perspective of a third qubit.  In this work we characterise the QCMI of tripartite W-states when some of the qubits
are under accelerated motion.  Here for our investigations we consider a massless fermionic field in the single mode approximation. 
We consider all possible situations with respect to acceleration of the qubits.  From our results we observe that QCMI can either 
increase or decrease depending on the role of the qubit being accelerated.  Finally we discuss the connection between QCMI and correlations 
by studying the biseparable and separable states.  
\end{abstract}

\noindent{\it Keywords\/}: Conditional mutual information, Quantum conditional mutual information, Non-inertial, W state, Single-mode approximation

\maketitle

\tableofcontents

\section{Introduction}
Quantum physics generalizes classical physics to systems in which the inherent wave-particle duality is explicitly observed. Typically, investigations on quantum systems are conducted within inertial frames where qubits are either stationary or under uniform motion \cite{Unruh1984inertial, Dowling2008inertial, Nicolai2010inertial, Lanzagobookinertial}.  When qubits undergo acceleration they are in a non-inertial 
frame of reference. Investigating the properties of qubits in non-inertial frames of reference is crucial to understanding the effects of acceleration 
on quantum systems \cite{ZhouQKD2018, Pierini2018}. Entanglement is a very common and well-known property of quantum systems and has been the subject of investigation in several works 
where non-inertial frames were considered \cite{Hwang2001entanglement, Fuentes2005entanglement, Fuentes2006singlemode, Adesso2007entanglement, 
Dehnavi2011entanglement, Hwang2012entanglement-singlemode, Ramzan2012entanglement, Wang2017entanglement, Jing2018entanglement,Dong2019wstate}. 
Similar investigations were also carried out on quantum coherence \cite{Baumgratz2014coherence, Wu2021coherence, Savee2022coherence, Wu2022coh}, 
quantum discord \cite{Animesh2009discord, Wang2010discord, Ramzan2014discordBSMA, Sugumi2016discord} and several other correlations 
\cite{Martinez2010corr, Martinez2010boscorr, Adesso2012corr, Dragan2013localprojective, Wang2016quantumsteering}.   

The correlation between two qubits can be characterised using the quantum mutual information \cite{Zurek2001discord,Vedral2001discord}. A related quantity is the quantum 
conditional mutual information (QCMI). The QCMI of a tripartite system is a measure of the correlation between two of the parties, given that 
information from third party is shared.  The QCMI of quantum systems has been studied in inertial frames in the context of quantum state reconstruction
\cite{Brandao2015redistribute}.  In fact QCMI gives the communication cost of transferring information between two parties \cite{Devtak2008exactcost}. 
Additionally, in \cite{Chandra2020CMI} it was shown that the QCMI can be used to measure quantum correlations in a tripartite system and consequently 
can be extended to measure correlations in quantum systems with arbitrary numbers of qubits. Since QCMI has several useful interpretations and applications 
it is essential to understand how it is affected by acceleration. In the present work we study the effect of acceleration on QCMI considering 
tripartite W-states, a class of symmetric, multiparty entangled states. We consider situations where either one or two of the qubits are under acceleration. 
The QCMI is an asymmetric function in the sense that it characterises the correlation between a pair of qubits from the perspective of the 
third qubit.  So we have a qubit carrying conditioning information and the correlation between the remaining qubits are measured from the 
knowledge of this conditioning information.  In the case of a single accelerated qubit, we can choose to accelerate either the qubit 
carrying the conditioning information or the qubits whose correlations we are measuring.  While studying the scenario where two qubits are simultaneously 
accelerated, we consider the following situations: (a) when one of the accelerated qubits is the one carrying conditioning information  and (b) when 
none of the accelerated qubits are the conditioning qubits.

In our present work we investigate the QCMI of tripartite $W$-states.  The $W$-states have an entanglement distribution which is quite different from that of the GHZ state.  In the GHZ state the entanglement is distributed in a more global manner where the loss of single 
qubit results in the loss of the total entanglement in the system.  Whereas in the $W$-state, the entanglement is distributed more locally 
where the loss of qubits leads to only a partial loss of entanglement in the system. To investigate the non-inertial effects in states with bipartite entanglement distribution we study the W states in our work. In Sec. \ref{fermionic} we provide an overview of 
relativity in quantum information for a fermionic field.  A brief introduction to the quantum version of the mutual information and 
conditional mutual information is given in Sec. \ref{QCMI}.  The QCMI of tripartite W-states when a single qubit is accelerated is given in 
Sec. \ref{RQCMI:1qubitacc}.  The calculation corresponding to two qubit acceleration scenario is given in Sec. \ref{RQCMI:2qubitacc}.  In 
section \ref{analysis} we analyze the results and also describe the QCMI of biseparable and separable states.  We present our conclusions in 
Section \ref{conclusion}.

\section{Fermionic field under uniform acceleration}
\label{fermionic}

The events corresponding to an inertial observer can be described using Minkowski coordinates $(x,y,z,t)$ in the 
Minkowski spacetime. When the observer is accelerated, their frame of reference becomes non-inertial and consequently we need to use Rindler co-ordinates here. In our work we consider a setting with one spatial and one time dimension $(z,t)$. 

The Minkowski spacetime is divided into four wedges, where the regions I and II represent the two causally disconnected Rindler regions \cite{Brushi2010BSMA}. The future and past light cones are represented by F and P and their corresponding event horizons are described by the equations $|z| = t$ \& $|z| = -t$ respectively. In the Minkowski coordinates the world lines of a uniformly accelerated observer is a hyperbola. In the Rindler 
coordinates, each region contains one branch of the hyperbola. The co-ordinates corresponding to these two different regions are: 
\begin{eqnarray}
 t &= a^{-1} e^{a \xi} \sinh a \tau,  \,\,   z = a^{-1} e^{a \xi} \cosh a \tau, \,\,\,\, |z| < t \nonumber \\
 t &= -a^{-1} e^ {a \xi} \sinh a \tau,  \,\, z = a^{-1} e^{a \xi} \cosh a \tau, |z| > t ~,
\label{coord}
\end{eqnarray}
where $a$ is the acceleration and $(\xi,\tau)$ are the space-like coordinate and the proper time respectively.
The massless Dirac field equation  is given by,
\begin{equation}
i(\gamma^{0}\partial_{0} + \gamma^{\bf 3}\partial_{\bf z})\psi = 0~,
\label{Dirac}
\end{equation}
where $\gamma^0$ and $\gamma^{\bf 3}$ are the Dirac-Pauli matrices and $\psi$ is the spinor wave function. The field can be 
described in the Minkowski coordinates for an inertial observer. It can be expanded in terms of the positive and negative energy solutions and they form a complete orthonormal set of modes,
\begin{equation}
\psi = \int dk (a_k \psi^+_k + b_k^\dagger \psi_k^-)~,
\label{psi_mink}
\end{equation}
where $\psi^+$ and $\psi^-$ are the particle and anti-particle wave functions. The creation and annihilation operators are $a_k^\dagger (b_k^\dagger)$ and $a_k (b_k)$ for the positive(negative) energy solution with momentum $k$. These operators satisfy the anti-commutation relations,
\begin{equation}
\{a_i, a_j^\dagger\} = \{b_i, b_j^\dagger\} = \delta_{ij}~,
\label{operator}
\end{equation}
with the other anti-commutators vanishing in Minkowski space-time. For the accelerated observer one can define their motion using Rindler coordinates in Minkowski space-time. The two regions I and II are causally disconnected from each other and so they have to be independently quantized. For the positive and negative energy 
solutions $\psi^{\mathrm{I}+}$, $\psi^{\mathrm{I}-}$ for region-I and $\psi^{\mathrm{II}+}$, $\psi^{\mathrm{II}-}$ for region-II, the fermionic fields have distinct creation and annihilation operators. In Rindler region-I for the particles i.e., positive energy solutions,
the creation and annihilation operators are $c_k^{\dagger \mathrm{I}}$, $c_k^\mathrm{I}$. Meanwhile, $d_k^{\dagger \mathrm{I}}$, $d_k^\mathrm{I}$ are the creation and annihilation operators for the antiparticles (negative energy solutions). Similarly for Rindler region-II, the $c_k^{\dagger \mathrm{II}}$, $c_k^\mathrm{II}$ are the creation and annihilation operators for the particles and $d_k^{\dagger_\mathrm{II}}$, $d_k^\mathrm{II}$ are 
the creation and annihilation operators for the antiparticles. The Dirac field is now written in terms of four solutions, where the first pair corresponds to the first Rindler region-I and the second pair for the second Rindler region-II. The Dirac equation in the Rindler coordinate is, 
\begin{equation}
\psi = \int dk(c_k^\mathrm{I} \psi^\mathrm{I+}_k + d_k^{\mathrm{I} \dagger} \psi^\mathrm{I-}_k + c_k^\mathrm{II} \psi^\mathrm{II+}_k + d_k^\mathrm{II \dagger}\psi^{\mathrm{II-}}_k)~.
\label{psi_rind}
\end{equation}
The operators in Eq.~(\ref{psi_rind}) will obey the usual fermionic anti-commutation relations 
$\{c_k^{i}, c_{k'}^{ i \dagger}\}=\{ d^{i}_k, d^{i \dagger}_{k'}\}=\delta_{kk'}^{i}$, where $i$ represents either Rindler 
region-I or -II. There is no anti-commutation relationship between the particle and the antiparticle operators. Also, the anti-commutation relation between any two operators from distinct causally disconnected regions is always zero. 
The creation and annihilation operators of the Minkowski and Rindler coordinates are related through a Bogoliubov transformation. In fact there are two sets of transformations as explained below: The first set of 
transformations relates the particle in region-I to the anti-particle in region-II,
\begin{equation}
a_k = \cos r c_k^\mathrm{I} - e^{-i\phi} \sin r  d_{-k}^{\mathrm{II} \dagger}; \quad
b_{-k}^{\dagger} = e^{i\phi} \sin r c_k^\mathrm{I} + \cos r d_{-k}^{\mathrm{II} \dagger}~.
\label{Bogtrans}
\end{equation}
In the second set of transformation equations, the anti-particle in region-I and the particle in region-II are used,
\begin{equation}
 b_k = \cos r d_k^\mathrm{I} - e^{-i\phi} \sin r  c_{-k}^{\mathrm{II} \dagger}; \quad
 a_{-k}^{\dagger} = -e^{-i\phi} \sin r d_k^\mathrm{I} + \cos r c_{-k}^{\mathrm{II} \dagger}~.
\label{BogtransI}
\end{equation} 
In Eqs.~(\ref{Bogtrans}) and (\ref{BogtransI}) the factor $r = \tan^{-1}e^{(- \pi \Omega)}$ is the acceleration parameter. Here $\Omega \equiv \omega c/a$ is the Rindler frequency where $\omega$ is the Minkowski frequency, $c$ is the 
velocity of light and $a$ is the proper acceleration of the system. The trivial phase factor $\phi$ can be absorbed into the 
creation and annihilation operators. A detailed account of the Bogoliubov transformations in the context of non-inertial quantum systems
can be found in Refs.~\cite{Soffel1980Bougli, Takagi-1986Bougli, Jaur1991Bougli}. This direct transformation from Minkowski coordinates to Rindler coordinates is valid only in 
the single mode approximation \cite{Fuentes2006singlemode}. In this approximation, Rob's (Relativistic Bob) detector observes a narrow and sharply  
peaked positive frequency $\Omega_{R}$ in Rindler region-I. This frequency is related to the single Minkowski positive frequency $\omega_{A}$ corresponding to inertial Alice.  The Minkowski and Rindler modes are related via the two-mode squeezing operator,
\begin{equation}
S = \exp[r(e^{-i \phi} c_k^{\mathrm{I} \dagger}d_{-k}^{\mathrm{II} \dagger} + e^{i \phi} c_k^\mathrm{I} d_{-k}^\mathrm{II} )]~.
\label{squeeze1}
\end{equation}
\begin{figure}
\centering
\includegraphics[width=0.5\columnwidth]{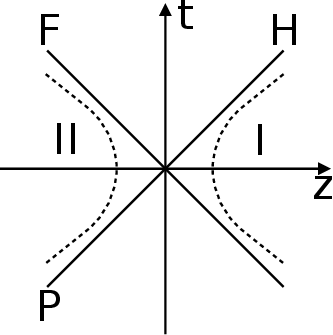}
\caption{The figure represents a space-time $(z,t)$ event consisting of two Rindler regions I and II. Here $P$ and $F$ are the past and future events and $H$ denotes the Rindler Horizon.}
\label{timeline}
\end{figure}
The relation between the creation and annihilation operators of the Minkowski and Rindler coordinates in terms of the squeezing operator is,
\begin{equation}
a_k =  S c_k^\mathrm{I}  S^\dagger;\qquad b_{-k}^\dagger = S d_{-k}^{\mathrm{II} \dagger} S^\dagger~.
\label{squeeze2}
\end{equation}
The Minkowski coordinates $(t,x)$ of the inertial observers and the Rindler coordinates $(\xi, \tau)$ of the accelerated observers can be related as \cite{Takagi-1986Bougli},
\begin{equation}
\xi = \sqrt{x^2 -t^2}~; \qquad
\tau = a \tanh (t/x)~,
\label{propertime}
\end{equation}
where $a$ is the proper acceleration. The range of the Rindler coordinates are $0< \xi < \infty$ and $ -\infty < \tau < + \infty$, where region-I has $|z| < t$ and for region-II we have $|z| > t$.

A vector field allows only two states for each mode. This is in contrast to the scalar field which allows an infinite mode decomposition. The Minkowski vacuum can be defined in terms of the Unruh vacuum. Here L and R are the left and right Unruh modes with frequency $\Omega$. Hence an element in the Fock basis 
is written in terms of left and right modes as \cite{Brushi2010BSMA},
\begin{equation}
|F_k \rangle = |F_k \rangle_R \otimes |F_k \rangle_L~,
\label{Fockbasis}
\end{equation}
with
\begin{equation}
|F_k \rangle_R = |n \rangle^+_\mathrm{I} |m \rangle^-_\mathrm{II}~, \qquad
|F_k \rangle_L = |p \rangle^-_\mathrm{I} |q \rangle^+_\mathrm{{II}}~,
\label{Fock_LR}
\end{equation}
where $\pm$ denotes the particle and antiparticle. 
To examine the case in a simpler manner we consider Grassmann scalars as they preserve the fundamental of Dirac field. Using the above ansatz, the Minkowski Fock vacuum state can be expressed in terms of the Rindler Fock states \cite{Brushi2010BSMA} as,
\begin{equation}
|0 \rangle_M  = \otimes_\Omega |0_\Omega \rangle_M = |0_\Omega \rangle_R \otimes |0_\Omega \rangle_L~, 
\label{Fock_LR_I_II}
\end{equation}
with the individual representations as follows: 
\begin{eqnarray}
|0_\Omega \rangle_R &= \cos r |0_\Omega \rangle^+_\mathrm{I} |0_\Omega \rangle^-_\mathrm{II} + \sin r |1_\Omega \rangle^+_\mathrm{I} |1_\Omega \rangle^-_\mathrm{II}, \nonumber \\
|0_\Omega \rangle_L &= \cos r |0_\Omega \rangle^-_\mathrm{I} |0_\Omega \rangle^+_\mathrm{II} - \sin r |1_\Omega \rangle^-_\mathrm{I} |1_\Omega \rangle^+_\mathrm{II}~. 
\label{Fock_expansion}
\end{eqnarray}
Substituting Eq.~(\ref{Fock_expansion}) into Eq.~(\ref{Fock_LR_I_II}) we obtain, 
\begin{equation}
\fl |0_\Omega \rangle_M = \cos^2 r_\Omega |0000 \rangle_\Omega - \sin r_\Omega \cos r_\Omega |0011 \rangle_\Omega + \sin r_\Omega \cos r_\Omega |1100 \rangle_\Omega - \sin^2 r_\Omega |1111\rangle_\Omega~,
\label{vacuumstate_BSA}
\end{equation}
where $\pm$ denotes the particle and antiparticle vacuum with momentum modes $\pm k$. We note that $a_k |0_k \rangle^+_M = b_{-k}|0_k \rangle^-_M = 0$ where $a_{k}$ and $b_{k}$ are the annihilation 
operators corresponding to the Minkowski particle and the antiparticle vacuum state.  Meanwhile, the action of the Rindler creation and annihilation operators on the Rindler vacuum state is,
\begin{eqnarray}
 c^I_k|0_k \rangle^+_\mathrm{I} =0; \qquad
d^\mathrm{II}_{-k}|0_{-k}\rangle^-_\mathrm{II} = 0, \nonumber\\
 c^{\mathrm{I} \dagger}_k|0_k \rangle^+\mathrm{I} = |1_k \rangle^+_\mathrm{I}; \qquad
d^{\mathrm{II} \dagger}_{-k}|0_{-k}\rangle^-_\mathrm{II} = |1_{-k} \rangle ^-_\mathrm{II}~.
\label{Rindler_vacuum}
\end{eqnarray}
The Minkowski single particle excited state can be written in terms of left and right Unruh operators. Here $q_R$ and $q_L$ are the
complex coefficients of the operators in the right (R) and left (L) region, satisfying $|q_R|^2 + |q_L|^2 = 1$. The single particle excited state is:
\begin{equation}
\fl|1_\Omega \rangle^+_U = q_R [\cos r_\Omega |1000\rangle_\Omega - \sin r_\Omega |1011\rangle_\Omega] 
+ q_L[\sin r_\Omega |1101 \rangle_\Omega + \cos r_\Omega |0001\rangle_\Omega]~,
\label{excitedstate_BSA}
\end{equation}
where $a_k ^\dagger |1 \rangle = 0$.  
The two modes $|0_k \rangle^+$ and $|1_k \rangle^-$ represent the positive frequency modes corresponding to Alice, the inertial observer. Since Rob is moving with uniform acceleration, there are two causally disconnected regions {\it viz} Rindler regions-I and -II. Rob has access to only one of these two regions while the other region is completely inaccessible. Hence, any quantum property described by a single Minkowski mode
is described jointly between the two Rindler modes. Due to the absence of a causal connection between these two modes there is a decoherence in the system. In this work we consider tripartite systems which are initially inertial and consider when either one or two of these particles undergo acceleration. 

Eqs.~(\ref{vacuumstate_BSA}) \& (\ref{excitedstate_BSA}) give the Minkowski vacuum and single particle excited state
beyond the single mode approximation \cite{Brushi2010BSMA}. Choosing $|q_R|=1$ and $|q_L|=0$ leads to the limiting case referred to as the single mode approximation. In this work we consider this regime. The vacuum and excited states  corresponding to the single mode approximation are \cite{Fuentes2006singlemode}:
\begin{eqnarray}
|0_k \rangle ^+_M &= \cos r |0_\Omega \rangle^+_\mathrm{I} |0_{-\Omega} \rangle^-_\mathrm{II} + e^{-i\phi}\sin r |1_{\Omega} \rangle^+_\mathrm{I} |1_{-\Omega}\rangle^-_\mathrm{II}, \nonumber \\
|1_k \rangle ^+_M %&= a^\dagger |0_k \rangle^+_M \nn \\ 
&= |1_{\Omega} \rangle^+_\mathrm{I} |0_{-\Omega}\rangle^-_\mathrm{II}~.
\label{SMA}
\end{eqnarray}
%
%%%%%%%%%%%%%%%%%%%%%%%%%%%%%%%%%%%%%%%%%%%%%%%%%%%%%%%%%%%%%%%%%%%%%%%%%%%%%
%
%                 Quantum Conditional mutual information
%
%%%%%%%%%%%%%%%%%%%%%%%%%%%%%%%%%%%%%%%%%%%%%%%%%%%%%%%%%%%%%%%%%%%%%%%%%%%%%
%

\section{Quantum conditional mutual information and quantum Markov chains} 
\label{QCMI}

\begin{figure}
\includegraphics[width=\columnwidth]{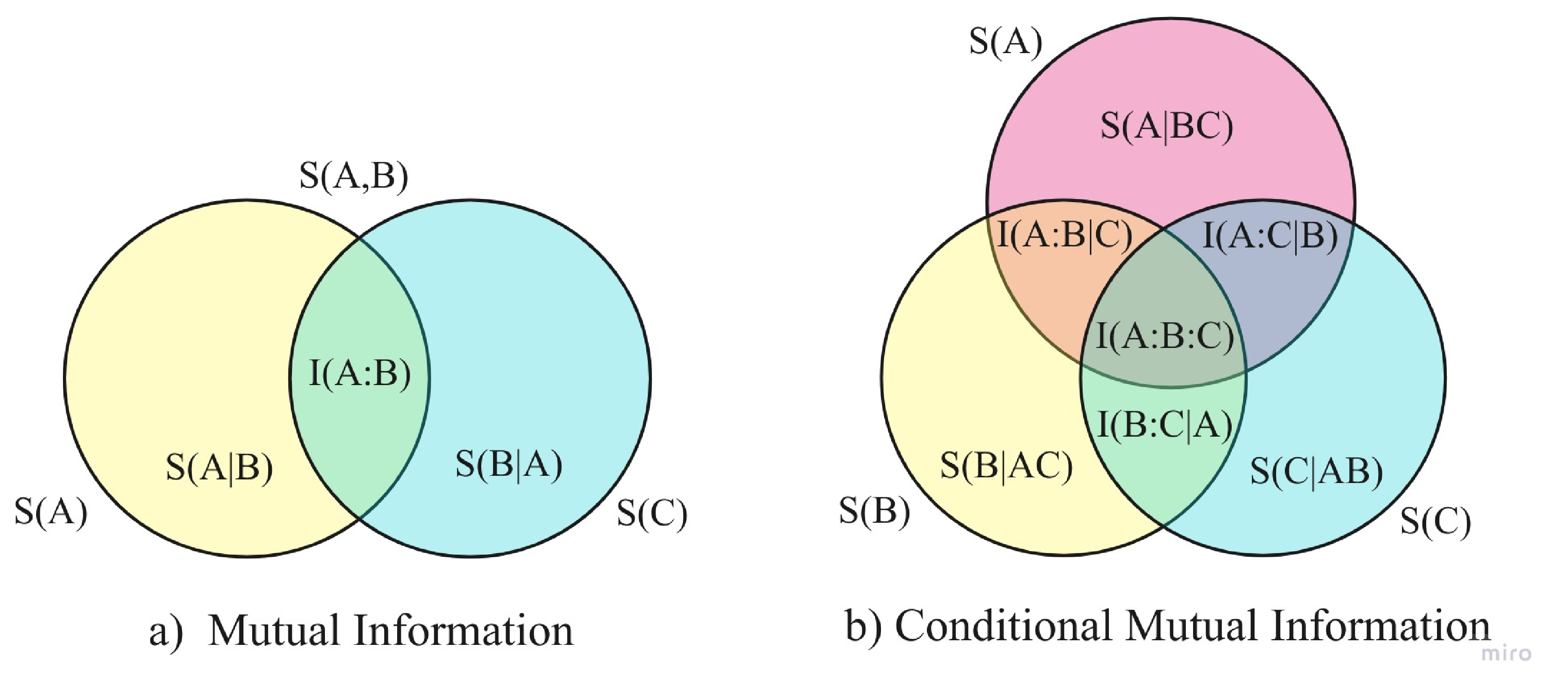}
\caption{The Venn diagram representation of mutual and conditional mutual information in terms of entropies for corresponding bipartite and tripartite systems.}
\label{venn_nonrel}
\end{figure}

The mutual information quantifies the amount of information shared in common between the different constituents of a composite system.  In classical information theory, the mutual information between two jointly distributed random variables $A$ and $B$ is defined as \cite{Thomas1991Inftheory},
\begin{equation}
I(A:B) =  S(A) + S(B) - S(AB)~,
\label{cmi}
\end{equation}
where $S(A)$ and $S(B)$ are the Shannon  entropies of the individual systems $A$ and $B$ and $S(AB)$ is the entropy 
of the joint system. In Fig.~(\ref{venn_nonrel}) we use a Venn diagram to illustrate these different quantities: entropies $S(A)$ and $S(B)$, joint entropy $S(AB)$, conditional entropy $S(A|B)$, and mutual information $I(A:B)$. If $A$ and $B$ are independent then $I(A:B)=0$. For a perfectly correlated pair of random variables $I(A:B)= \log_2 d$ where $d$ is the dimension of the Hilbert space. For a bipartite quantum system $\rho_{AB}$, the quantum mutual information (QMI) \cite{Bennett1999QMI,Jaeger2006Quantuminformation} is,
\begin{equation}
I(A:B) = S(\rho_{A}) + S(\rho_{B}) - S(\rho_{AB})~,
\label{qcmi}
\end{equation}
where $S(\rho_{A})$ is the von Neumann entropy of the reduced system $\rho_{A} = {\rm tr}_{B} (\rho_{AB})$ and $S(\rho_{AB})$ is the entropy of the joint system. The von Neumann entropy in terms of density operator is defined as 
\begin{equation}
 S(\rho) = -tr (\rho \ln \rho) = - \sum_i \lambda_i \log \lambda_i~, 
\end{equation}
where $\lambda_i$ are the eigenvalues of $\rho$.
The quantum mutual information characterises the amount of information shared between two quantum parties. Hence, the quantum mutual information is a measure of the total correlation present in a bipartite system. For a bipartite system the range of QMI is
\begin{equation}
0 \leq I(A:B) \leq \log_2 d~.
\label{biparcond}
\end{equation}
In a correlated tripartite system, an interesting information theoretic quantity is the conditional mutual information (CMI) \cite{Thomas1991Inftheory}. To define CMI let us consider three different jointly distributed discrete random variables $A$, $B$ and $C$. Here the conditional mutual information is the expected value of two random variables given the value of the third, 
\begin{equation}
I(A:B|C) = S(AC) + S(BC) - S(ABC)-S(C)~,
\label{cmi3}
\end{equation}
where $S(X)$ is the Shannon entropy of $X$. A schematic picture of three jointly distributed discrete random variables is shown in Fig.~(\ref{venn_nonrel}{\color{red}{b}}) using a Venn diagram. From the picture we can understand the various quantities $I(A:B|C)$, $I(A:C|B)$, $I(B:C|A)$ and $I(A:B:C)$. When $A$ and $B$ are independent, 
\begin{equation}
 I(A:B|C)=0~.
\end{equation}
Quantum conditional mutual information is maximised when $A$ and $B$ are perfectly correlated and $C$ is disjoint (orange shaded portion in the Venn diagram of Fig.~(\ref{venn_nonrel}), and so $I(A:B|C) = I(A:B)$).  When the qubits are accelerated, the role 
of the accelerated qubit as to whether it is the correlated qubit or the conditional qubit decides whether the QCMI in the system increases
or decreases. This is discussed in detail in section {\ref{analysis}}.

The quantum version of the conditional mutual information for a tripartite system $\rho_{ABC}$ is,
\begin{equation}
I(A:B|C) = S(\rho_{AC}) + S(\rho_{BC}) - S(\rho_{ABC})-S(\rho_{C})~.
\label{QCMI3}
\end{equation}
Here $S(\rho_{ABC})$ is the von Neumann entropy of the tripartite system. The quantities $S(\rho_{AC})$, $S(\rho_{BC})$ and $S(\rho_{C})$ are the von Neumann entropies of the respective reduced density matrices. The quantum conditional mutual information (QCMI) of a tripartite state $\rho_{ABC}$ on $\mathcal{H}_{A} \otimes \mathcal{H}_{B} \otimes \mathcal{H}_{C}$ quantifies the correlations between $A$ and $B$ from the point of view of $C$. 

The QCMI of a tripartite quantum system is bounded as $0 \leq I(A:B|C) \leq \log_2 d$. The conditional mutual information can be used to characterise a Markov chain \cite{Omar2015qcmiMarkov,Huang2021nonmarkov}. A connection between QCMI and reconstruction of states was first given by Fawzi-Renner inequality \cite{Omar2015qcmiMarkov} which is defined as follows:
$$
I(A:B|C)_{\rho} \geq \min_{\Lambda:C\rightarrow CA} S_{1/2} [ \rho_{CAB}||\Lambda \otimes id_R(\rho_{CB})]~.
$$

 In a Markov chain $A-C-B$, one can recover the lost input $A$, purely from the quantity $C$. Thus, a Markov chain implies recoverability of information and hence $I(A:B|C)=0$. A quantum Markov chain is a tripartite density matrix $\rho_{ABC}$ in which we can reconstruct system $B$ through the action of a map on system $C$. This map $\tau_{C\rightarrow BC}$ is a quantum operation from $C$ to $B \otimes C$, called a recovery map which acts such that $\rho_{ABC} = \tau_{C\rightarrow BC} (\rho_{AC})$ \cite{Omar2016recoverymap}. When $I(A:B|C) = 0$, the state $\rho_{ABC}$ can be exactly reconstructed. For nonzero values of QCMI, the tripartite state $\rho_{ABC}$ is not exactly reconstructed and $I(A:B|C)$ gives an upper bound on its distance to the closest reconstructed state.  
 
In the present work, we consider tripartite correlated quantum states. We consider the situations where either one or two qubits are under acceleration. Under these conditions we compute QCMI of these tripartite quantum states and analyse quantum state reconstruction in non-inertial frames of references. 

\section{Relativistic quantum conditional mutual information: single qubit acceleration} 
\label{RQCMI:1qubitacc}

The quantum conditional mutual information (QCMI) of a three qubit system is discussed when one of the qubits is subjected to acceleration. From the expression of QCMI in 
Eq.~(\ref{QCMI3}) one can see that it is an asymmetric function. We are measuring the correlations between qubits $A$ and $B$ conditional on the information in qubit $C$. Hence the 
role played by qubits $A$ and $B$ is quite different from the role played by qubit $C$. In this work we examine the effect of acceleration when either qubit $B$ or $C$ is 
accelerated. A tripartite qubit system can be correlated either in a genuinely tripartite way or in a bipartite fashion. A maximally-entangled tripartite system like a GHZ state 
looses all its quantum correlations when any qubit is traced out. When the qubits are correlated in a bipartite fashion as in a $W$-state, the loss of a single qubit might not 
result in the complete loss of quantum correlations in the system. In the tripartite system, the three parties Alice, Bob and Charlie have monochromatic detectors with 
$\omega_{A}$, $\omega_{B}$ and $\omega_{C}$ respectively. Since our investigations are based on single mode approximations we consider $\omega_{A} \sim \omega_{B} \sim \omega_{C} 
\approx \omega$.  

\subsection{Acceleration of Charlie's qubit }

A three qubit $W$-state which is correlated in a bipartite manner has the following form:  
\begin{equation}
|W \rangle = \frac{1}{\sqrt{3}} (|0_{A} 0_{B} 1_{C} \rangle  + |0_{A} 1_{B} 0_{C} \rangle + |1_{A} 0_{B} 0_{C})~, 
\label{3w}
\end{equation}
where $A$, $B$ and $C$ refer to the three parties Alice, Bob and Charlie. Under the condition where Charlie undergoes uniform acceleration, the state of his qubit in the Minkowski vacuum can be replaced by the corresponding Rindler modes. 
Hence the $W$-states when Charlie is under non-inertial motion can be written as,
\begin{eqnarray}
\fl |W \rangle &= \frac{1}{\sqrt{3}}\bigg( |00 \rangle |1_{k} \rangle^{+}_\mathrm{I} |0_{-k} \rangle^{-}_\mathrm{II} + \cos r (|01 \rangle  |0_{k} \rangle^{+}_\mathrm{I}  |0_{-k} \rangle^{-}_\mathrm{II} + |10 \rangle   |0_{k} \rangle_\mathrm{I}^{+}  |0_{-k} \rangle_\mathrm{II}^{-}) \nonumber \\
\fl &+ e^{-i\phi}\sin r \big(| 01 \rangle  |1_{k} \rangle_\mathrm{I}^{+}  |1_{-k} \rangle_\mathrm{II}^{-} 
+ |10 \rangle   |1_{k} \rangle_\mathrm{I}^{+}  |1_{-k} \rangle_\mathrm{II}^{-}\big)\bigg)~,
\label{w_caccel}
\end{eqnarray}
and the corresponding density matrix is,
\begin{eqnarray}
\fl \rho_{W}  &= \frac{1}{3} \bigg(\cos^2 r(|0100 \rangle \langle 0100|+ |0100 \rangle \langle 1000| 
+ |1000 \rangle \langle 1000| ) +|0010 \rangle \langle 0010|  \nonumber \\ 
\fl &+ \cos r \left(|0010 \rangle \langle 0100| + |0010 \rangle \langle 1000| \right)
+ \sin^2 r( |0111 \rangle \langle 0111|+|0111 \rangle \langle 1011| \nonumber \\
\fl &+ |1011 \rangle \langle 1011|) + e^{i\phi} \sin r \big( |0010\rangle \langle 0111| 
+ |0010 \rangle \langle 1011|  +  \cos r ( |0100\rangle \langle 0111|  \nonumber \\ 
\fl &+ |0100 \rangle \langle 1011|  +  |1000 \rangle \langle 0111| + |1000 \rangle \langle 1011 |)\big) + (h.c.)_{od} \bigg)~,
\label{w_caccel_dm}
\end{eqnarray}
where $h.c.$  and $od$ refers to hermitian conjugate and  off-diagonal elements of the density matrix respectively. 
The accelerated qubit of Charlie has two modes corresponding to the two causally disconnected Rindler regions. Here 
we retain only the Rindler region-I in our studies and the resulting expression upon tracing our region-II is,
\begin{eqnarray}
\fl \rho_{W}  &= \frac{1}{3} \bigg( \cos^2 r (|010 \rangle \langle 010|+ |010 \rangle \langle 100| + |100 \rangle \langle 100| )
+ \cos r \left(|001 \rangle \langle 010|+ |001 \rangle \langle 100| \right) \nonumber \\ 
\fl &+ |001 \rangle \langle 001|+\sin^2 r ( |011 \rangle \langle 011| + |011 \rangle \langle 101|+ |101 \rangle \langle 101|)+ (h.c.)_{od} \bigg)~.
\label{w_accel_dmtr}
\end{eqnarray}
The reduced density matrices $\rho_{AC}$, $\rho_{BC}$ and $\rho_{C}$ needed for calculating the quantum conditional mutual information are,
\begin{eqnarray}
\fl \rho_{AC} = \rho_{BC} &= \frac{1}{3}\bigg( \cos^2 r(|00 \rangle \langle 00 | + |10 \rangle \langle 10|)
+ \cos r|01 \rangle \langle 10 | + (1+\sin^{2} r) |01 \rangle \langle 01 | \nonumber \\
\fl &+ \sin^2 r |11 \rangle \langle 11| + (h.c)_{nd} \bigg), \nonumber \\
\fl &\rho_{C} = \frac{1}{3}(2 
\cos^{2} r  |0 \rangle \langle 0|+(1 + 2 \sin^{2} r)  |1 \rangle \langle 1 | )~. 
\label{w_caccel_rdm}
\end{eqnarray}
The corresponding eigenvalues are:
\begin{eqnarray*}
\fl \lambda_{ABC} = \left\{\frac{1-\cos2r}{3}, \frac{2+\cos2r}{3}\right\}~; \\
\fl \lambda_{AC} = \lambda_{BC} = \left\{ \frac{\cos^2 r}{3}, \frac{4 \pm \sqrt{14+2\cos4 r}}{12}, \frac{\sin^2 r}{3}\right\}~; \\
\fl \lambda_C = \left\{\frac{1+2\sin^2r}{3}, \frac{2\cos^2r}{3}\right\}~.
\end{eqnarray*}
The entropies of the $W$-state and its reduced density matrices are given by,
\begin{eqnarray}
\fl S(\rho_{ABC}) &= -\left(\frac{1+2 \cos^2 r}{3}\right) \log_2 \left(\frac{1+2 \cos^2 r}{3} \right)
- \left(\frac{2 \sin^2 r}{3}\right)\log_2 \left(\frac{2 \sin^2 r}{3}\right), \nonumber \\          
\fl S(\rho_{AC}) &= S (\rho_{BC}) 
= -\left(\frac{\sin^2 r}{3}\right) \log_2  \left(\frac{\sin^2 r}{3} \right)  - \left(\frac{\cos^2 r}{3}\right) \log_2 \left(\frac{\cos^2 r}{3} \right)\nonumber\\
\fl &+\frac{2}{3} \log_2 \left(\frac{\sin r \cos r}{3}\right)
- \frac{\sqrt{3+ \cos^2 2r}}{6}\log_2 \left(\frac{2- \sqrt{3+ \cos^2 2r}}{\sin 2 r}\right),\nonumber\\
\fl S(\rho_{C})  &= -\left(\frac{2\cos^2 r}{3}\right) \log_2 \left(\frac{2 \cos^2 r}{3}\right)
-\left(\frac{1+2 \sin^2 r}{3}\right) \log_2 \left( \frac{1+2 \sin^2 r}{3} \right)~,  
\label{w_caccel_entropy}
\end{eqnarray}
from which we find the quantum conditional mutual information to be,
\begin{eqnarray}
\fl I^C_W &= \frac{2}{3}-\frac{4}{3} \log_2 \left(\frac{\cos r \sin r}{3}\right) 
- \frac{2\sqrt{3+\cos^2 2r}}{3} \log_2 \left( \frac{2-\sqrt{3+\cos^2 2r}}{\sin 2r} \right) \nonumber \\
\fl &+ \left(\frac{1+2 \cos^2 r }{3}\right)\log_2 \left( \frac{1+2\cos^2 r}{3} \right)
+ \left(\frac{1+2\sin^2 r}{3}\right) \log_2 \left( \frac{1+2\sin^2 r}{3} \right).
\label{w_c_acc_qcmi}
\end{eqnarray}
\begin{figure*}[!htb]
 \includegraphics[width=\linewidth]{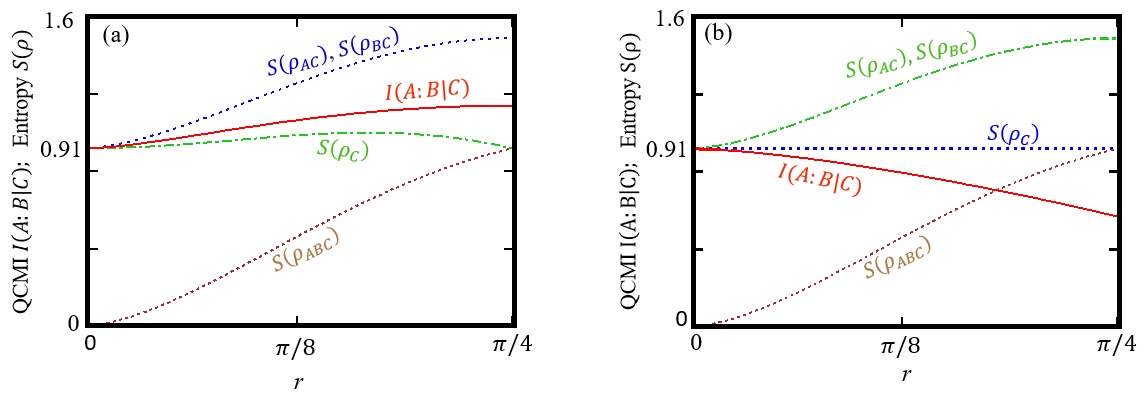}
\caption{ The variation of QCMI $(I)$, entropies of three qubit $(S(\rho_{ABC}))$ system and two $(S(\rho_{AC}), S(\rho_{BC}))$ and one qubit $(S(\rho_C))$ reduced systems as a function of acceleration parameter r is given for $W$ state under (a) C acceleration and (b) B acceleration. The variables $r$, QCMI and Entropy are dimensionless quantities.}
\label{1qubitaccel}
\end{figure*}

For $r=0$ in Eq.~(\ref{w_c_acc_qcmi}) we will obtain the QCMI of the W-state corresponding to the situation where 
all three parties are in an inertial frame. From Fig.~(\ref{1qubitaccel}{\color{red}{a}}) as Charlie accelerates, the QCMI increases steadily and attains its maximal value for $r \rightarrow \pi/4$ , the limit of infinite acceleration \cite{Fuentes2006singlemode}. The accelerated motion of Charlie leads to partial loss of information in the quantum state held by him. Due to this, the QCMI of the $W$-state increases implying that the quantum state reconstruction is harder when Charlie's qubit is under acceleration. 

\subsection{Acceleration of Bob's qubit}
The quantum conditional mutual information is an asymmetric function with respect to the role of Alice, Bob and Charlie. In QCMI we are measuring the correlation between qubits $A$ and $B$ given the information in qubit $C$, which is shared by $A$ and $B$. In the previous section we examined the situation when Charlie was under accelerated motion.  The motivation in the present section is to look at the effect of acceleration on QCMI when either $A$ or $B$ is under motion.
Next we consider a three qubit $W$-state in which Bob undergoes uniform acceleration. Under this condition, the 
$W$-state is transformed as, 
\begin{eqnarray}
\fl|W \rangle &= \frac{1}{\sqrt{3}} \bigg( \cos r (|0 \rangle |0_{k} \rangle^{+}_\mathrm{I} |0_{-k} \rangle^{-}_\mathrm{II} |1 \rangle + |1 \rangle  |0_{k} \rangle^{+}_\mathrm{I}  |0_{-k} \rangle^{-}_\mathrm{II} |0 \rangle)
+ |0 \rangle   |1_{k} \rangle_\mathrm{I}^{+}  |0_{-k} \rangle_\mathrm{II}^{-} | 0 \rangle \nonumber \\
\fl &+ e^{-i\phi}\sin r ( | 0 \rangle  |1_{k} \rangle_\mathrm{I}^{+}  |1_{-k} \rangle_\mathrm{II}^{-} |1 \rangle 
+ |1 \rangle |1_{k} \rangle_\mathrm{I}^{+}  |1_{-k} \rangle_\mathrm{II}^{-} |0 \rangle)\bigg)~.
\label{w_baccel}
\end{eqnarray}
The quantum state in Eq.~(\ref{w_baccel}) is being expressed using Rindler coordinates, where we have two 
causally disconnected Rindler modes, one each in Rindler regions-I and -II. Since these two modes are causally disconnected, we can trace out the mode in Rindler region-II and the resulting density matrix is given by,
\begin{eqnarray}
\fl \rho_{W} &= \frac{1}{3} \bigg(\cos^2 r (|001 \rangle \langle 001| + |001 \rangle \langle 100| + |100 \rangle \langle 100|)
+ \cos r( |001\rangle \langle 010|  +|010\rangle \langle 100|) \nonumber \\
\fl &+ |010 \rangle \langle 010| + \sin^2 r (|011 \rangle \langle 011| + |011 \rangle \langle 110|  + |110 \rangle \langle 110|) +  (h.c.)_{od} \bigg)~. 
 \label{w_baccel_dmtr}
\end{eqnarray}
The eigenvalues corresponding to the density matrices are:
\begin{eqnarray*}
\fl \lambda_{ABC} = \left\{\frac{1-\cos2r}{3}, \frac{2+\cos2r}{3}\right\}~;\\
\fl \lambda_{AC} = \lambda_C = \left\{\frac{2}{3}, \frac{1}{3}\right\}~; \\
\fl \lambda_{BC} = \left\{\frac{\cos^2 r}{3}, \frac{4\pm \sqrt{14+2\cos4 r}}{12}, \frac{\sin^2 r}{3}\right\}~. 
\end{eqnarray*}
From the density matrix in Eq.~(\ref{w_baccel_dmtr}), we can calculate the reduced density matrices $\rho_{AC}$, $\rho_{BC}$ and $\rho_{C}$ and compute their corresponding entropies,
\begin{eqnarray}
\fl S(\rho_{ABC}) &= - \left(\frac{2 \sin^2 r}{3} \right) \log_2 \left(\frac{2\sin^2 r}{3} \right)
- \left(\frac{1+ 2\cos^2 r}{3}\right) \log_2 \left(\frac{1+ 2\cos^2 r}{3} \right), \nonumber \\
\fl S (\rho_{AC}) &= S(\rho_{c}) =\log_2 3 -\frac{2}{3}, \nonumber \\
\fl S (\rho_{BC}) &=  \left(\frac{\sqrt{3+\cos^2 2r}}{3}\right) \log_2 \left(\frac{2-\sqrt{3+\cos^2 2r}}{\sin2r}\right) 
-\left(\frac{1+\cos^2 r}{3}\right) \log_2 \left(\frac{\cos^2 r}{3}\right)\nonumber\\
\fl \hspace{1.25cm} &-\left(\frac{1+\sin^2 r}{3}\right) \log_2 \left(\frac{\sin^2 r}{3}\right)~.
\label{w_baccel_rdm}
\end{eqnarray}
Using the entropies we find the QCMI for the W-state to be,
\begin{eqnarray}
\fl I^B_W &= \left( \frac{\sqrt{3+\cos^2 2r}}{3} \right) \log_2 \left( \frac{2-\sqrt{3+\cos^2 2r}}{\sin2r} \right) 
- \left( \frac{1+\cos^2 r}{3} \right) \log_2 \left(\frac{\cos^2 r}{3}\right) \nonumber \\
\fl &+ \left( \frac{2\sin^2 r- \cos^2 r}{3}\right) \log_2 \left(\frac{\sin^2 r}{3}\right)
+ \left( \frac{1+2\cos^2 r}{3} \right) \log_2 \left( \frac{1+2\cos^2 r}{3} \right)~.
\label{w_baccel_entropy}
\end{eqnarray}
From Fig.~(\ref{1qubitaccel}{\color{red}{b}}), we find that the QCMI for the $W$-state decreases with acceleration. This is due to the decrease in the correlations between qubits $A$ and $B$, when the information in qubit $C$ remains constant.
A detailed discussion on the variation of QCMI with the acceleration parameter `$r$' is given in section \ref{analysis}, 
where we clearly explain the relation between the choice of the qubit and the QCMI features. 
\section{Relativistic quantum conditional mutual information: two qubit acceleration} 
\label{RQCMI:2qubitacc}

In the current section, we investigate the situation where two qubits of a tripartite system are under acceleration. %Also we consider both $GHZ$ and $W$ type tripartite states. 
Due to the asymmetric nature of QCMI  $I(A:B|C)$, there are two possible ways of choosing two qubits in a tripartite systems. These two scenarios are:
\begin{itemize}
\item One of the accelerated qubit to be the conditioning qubit $C$ and the other is the qubit held by Bob.
\item Alice and Bob hold the two accelerated qubits and the conditioning qubit held by Charlie is not accelerated.
\end{itemize}  
We examine both situations below.

\subsection{Acceleration of the qubits held by Bob and Charlie}

The tripartite $W$-state when Bob and Charlie are under acceleration is, 
\begin{eqnarray}
\fl|W \rangle^{BC} &= \frac{1}{\sqrt{3}}  \bigg( \cos r_1 | 0 \rangle  |0_{k} \rangle_\mathrm{I}^{+}  |0_{-k} \rangle_\mathrm{II}^{-} |1_{k} \rangle_\mathrm{I}^{+}  |0_{-k} \rangle_\mathrm{II}^{-}
+ \cos r_2 \big(|0 \rangle |1_{k} \rangle_\mathrm{I}^{+} |0_{-k} \rangle_\mathrm{II}^{-}
|0_{k} \rangle_\mathrm{I}^{+} |0_{-k} \rangle_\mathrm{II}^{-} \nonumber \\
\fl &+ \cos r_1 |1 \rangle |0_{k} \rangle_\mathrm{I}^{+}  |0_{-k} \rangle_\mathrm{II}^{-}|0_{k} \rangle_\mathrm{I}^{+}  |0_{-k} \rangle_\mathrm{II}^{-} \big)+ e^{-2i \phi} \sin r_1 \sin r_2 |1 \rangle |1_{k}\rangle _\mathrm{I}^{+} |1_{-k}\rangle_\mathrm{II}^{-}|1_{k}\rangle_\mathrm{I}^{+} 1_{-k}\rangle_\mathrm{II}^{-} \rangle \nonumber \\
\fl &+ e^{-i \phi} \big(\sin r_2 \cos r_1 |1 \rangle |0_{k} \rangle_\mathrm{I}^{+}  |0_{-k} \rangle_\mathrm{II}^{-}|1_{k} \rangle_\mathrm{I}^{+}  |1_{-k} \rangle_\mathrm{II}^{-} + \sin r_1 \cos r_2 |1 \rangle |1_{k} \rangle_\mathrm{I}^{+}  |1_{-k} \rangle_\mathrm{II}^{-}|0_{k} \rangle_\mathrm{I}^{+}  |0_{-k} \rangle_\mathrm{II}^{-}\nonumber \\
\fl &+ \sin r_1 | 0 \rangle  |1_{k} \rangle_\mathrm{I}^{+}  |1_{-k} \rangle_\mathrm{II}^{-} |1_{k} \rangle_\mathrm{I}^{+}  |0_{-k} \rangle_\mathrm{II}^{-} 
+ \sin r_2 |0 \rangle |1_{k} \rangle_\mathrm{I}^+ |0_{-k} \rangle_\mathrm{II}^{-} |1_{k} \rangle_\mathrm{I}^+ |1_{-k} \rangle_\mathrm{II}^- \big) \bigg)~.
\label{w_bc_accel}
\end{eqnarray}
The eigenvalues for the corresponding density matrices are:  
\begin{eqnarray}
\fl \lambda_{ABC} &= \Bigg\{\frac{\sin^2r_1\sin^2r_2}{3}~, \frac{5+3\cos2r_2+\cos2r_1(3+\cos2r_2)}{12}~, \nonumber \\
\fl &\frac{6-2(\cos2r_1-\cos2r_2)-\cos2(r_1 - r_2) -\cos2(r_1 + r_2)}{24}\nonumber \\
\fl &\pm \frac{\sqrt{32+8\cos4 r_1-32\cos2 r_1 \cos^2r_2-16\cos2r_2+\cos4 r_2}}{24}\Bigg\}~, \nonumber \\
\fl \lambda_{AC} &= \left\{\frac{\cos^2r_2}{3},\frac{4\pm\sqrt{14+2\cos4r_2}}{12}, \frac{\sin^2r_2}{3}\right\}~;\nonumber \\
\fl \lambda_C &= \left\{\frac{2\cos^2 r_2}{3}, \frac{1+2\sin^2r_2}{3}\right\}~;\nonumber\\
\fl \lambda_{BC} &= \Bigg\{\frac{\cos^2r_1 \cos^2r_2}{3},
\frac{\sin^2r_1+\sin^2r_2+\sin^2r_1\sin^2r_2}{3},\nonumber \\ 
\fl &\frac{6+2(\cos2r_1+\cos2r_2)-\cos2(r_1-r_2)-\cos2(r_1+r_2)}{24}\nonumber  \\
\fl &\pm\frac{\sqrt{32 + 8(\cos4r_1+\cos4r_2)+16\cos2r_2+32 \cos2 r_1 \sin^2 r_2}}{24}\Bigg\}~. 
\end{eqnarray}
Adopting the procedure of tracing out the Rindler II mode, we find the entropies of the total system as well as those of the 
subsystems. From these we find the quantum conditional mutual information of the $W$-state to be,
\begin{eqnarray}
\fl I^{BC}_W &=\frac{1}{6}(2\sin^2 r_2 + 2\sin^2 r_1 \cos^2 r_2- \xi_1) \log_2\frac{1}{6}(2\sin^2 r_2 + 2\sin^2 r_1 \cos^2 r_2- \xi_1)\nonumber \\
\fl &+\frac{1}{6}(2\sin^2 r_2 + 2\sin^2 r_1 \cos^2 r_2 + \xi_1) \log_2\frac{1}{6}(2\sin^2 r_2 + 2\sin^2 r_1 \cos^2 r_2 + \xi_1 )\nonumber \\
\fl &- \frac{1}{6}\big(2-2\sin^2 r_1 \sin^2 r_2 - \xi_3 \big) \log_2 \frac{1}{6} \big( 2-2\sin^2 r_1 \sin^2 r_2 - \xi_3 \big) + \frac{1}{12} \xi_4 \log_2 \xi_4 \nonumber \\ 
\fl &- \frac{1}{12}\left( 3+ 2\cos2r_1 \sin^2r_2 +\xi_2 \right) \log_2\frac{1}{12}\left( 3+ 2\cos2r_1 \sin^2 r_2 + \xi_2 \right) \nonumber \\
\fl &+\frac{\cos^2 r_2}{3}\big(2+\sin^2 r_1 \log_2\frac{\cos^2 r_2}{3}-\cos^2 r_1 \log_2 \cos^2 r_1\big) 
\nonumber \\
\fl &+ \left(\frac{1+2\sin^2 r_2}{3} \right)\log_2 \left( \frac{1+2\sin^2 r_2}{3}\right) + \left(\frac{\xi_6}{3}\right) \log_2 \left(\frac{3 \sin^2 r_1}{\xi_5}\right) \nonumber \\ 
\fl &- \left( \frac{\sqrt{1-\xi_7}}{3} \right) \log_2\left(\frac{1+\sqrt{1-\xi_7}}{1-\sqrt{1-\xi_7}} \right)- \frac{2}{3} \log_2 \left( \frac{\sin r \cos r}{3} \right) \nonumber \\
\fl &- \left( \frac{\sin^2 r_1 + \sin^2 r_2}{3} \right) 
\log_2 \left( \frac{\xi_5}{3} \right)- \frac{\cos^2 r_1 \sin^2 r_2}{3} \log_2\left(\frac{\sin^2 r_2}{3}\right)~,
\end{eqnarray}
where, 
\begin{eqnarray}
\fl \xi_1 &= \sqrt{2+\cos(2r_1+2r_2)\cos(2r_1-2r_2)-\cos2r_2-2\cos2r_1\cos^2r_2}, \nonumber \\
\fl \xi_2 &= \cos2r_2 + \sqrt{12+2\cos4r_1+2\cos4r_2-16\sin^2r_1\sin^2r_2},\nonumber \\
\fl \xi_3 &= \sqrt{\cos^2 2r_1 + \cos^2 2r_2 + 2 + 4 \sin^2 r_1 \sin^2 r_2}, \nonumber \\
\fl \xi_4 &= 5+6\cos(r_1+r_2)\cos(r_1-r_2) + \cos2r_2\cos2r_1,\nonumber \\
\fl \xi_5 &= \sin^2 r_1 (1+\sin^2 r_2) + \sin^2 r_2,\nonumber \\
\fl \xi_6 & = \sin^2 r_1 \sin^2 r_2,\nonumber\\
\fl \xi_7 &= \sin^2 r_2 \cos^2 r_2~.
\end{eqnarray}
In the limiting case of $r_1 = r_2 = r$, when both Bob ($r_1$) and Charlie ($r_2$) are uniformly accelerated the QCMI reduces to,
\begin{eqnarray}
\fl I^{BC}_W &= \frac{\cos^2 r}{3} \left(2 \sin^2 r +\cos^2 r \log_2 (2+\cos^2 r) + \log_2 \frac{(2+\cos^2 r)^4}{(2+\sin^2 r)^2}\right) \nonumber \\
\fl &+ \frac{\sin^2 r}{3} \bigg( \cos^2 r \log_2 \left( \frac{\sin^2 r}{2+\sin^2 r} \right) 
+ \sin^2 r \log_2 \left(\frac{{3\sin^4 r}}{{ 2\sin^2 r + \sin^4 r}}\right) \nonumber \\
\fl &+ \log_2 (\sin r +2\sin^3 r)^2 \left( \frac{6+3\cos^2 r}{2\sin^2 r + \sin^4 r}\right)^2 \bigg)+ 
\frac{1}{3} \log_2 (1+2\sin^2 r)\nonumber\\
\fl &+ \frac{\sqrt{3+\cos^2 2 r}}{3} 
\log_2 \left(\frac{ 2-\sqrt{3+\cos^2 2r}}{\sin 2r} \right)-\frac{7}{3} - \frac{2}{3} \log_2 \frac{\sin r \cos r}{3}~.
\label{W_bc_qcmi}
\end{eqnarray}
\begin{figure*}[!thb]
\includegraphics[width=\linewidth]{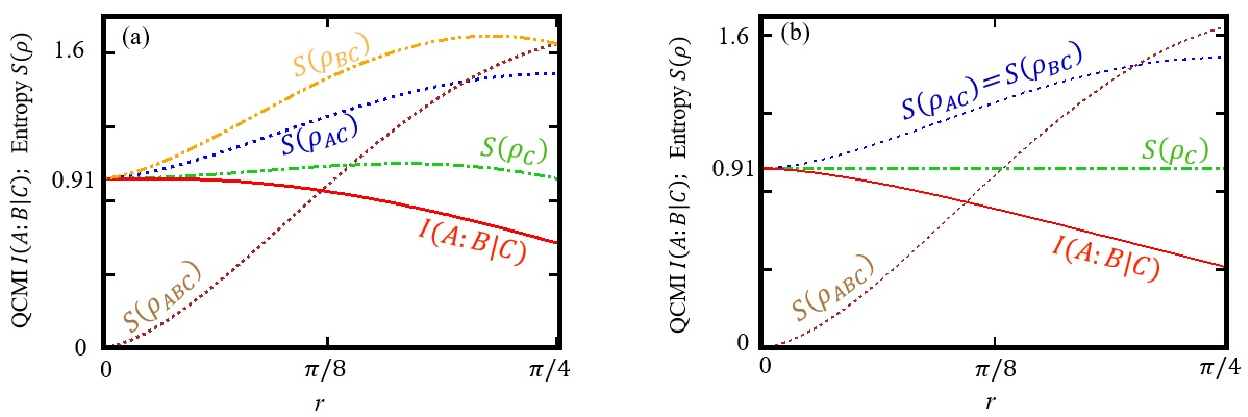}
\caption{The variation of QCMI $(I)$, entropies of three qubit $(S(\rho_{ABC}))$ system and two $(S(\rho_{AC}), S(\rho_{BC}))$ and one qubit $(S(\rho_C))$ reduced systems as a function of acceleration parameter r is given for $W$ state under (a) BC acceleration and (b) AB acceleration. The variables $r$, QCMI and entropy are dimensionless quantities.}
\label{2qubitaccel}
\end{figure*}
The QCMI of $W$-state falls steadily with acceleration.  In the single qubits scenario we find that QCMI increases with acceleration 
of qubit `$C$'.  Contrarily the QCMI decreases with acceleration of qubit `$B$'.  When both the qubits `$B$' and `$C$' are accelerated
the overall effect is a decrease in QCMI as observed in Fig.~(\ref{2qubitaccel}{\color{red}{a}}) .  

\subsection{Acceleration of the qubits used in mutual information (A and B)}
In this subsection we examine the situation where the qubits held by Alice and Bob are under acceleration. Here we measure the correlations between the accelerated qubits from the perspective of Charlie who is at rest. 

%\subsubsection{W-states}

In the three qubit $W$-state, when Alice and Bob undergo non-inertial motion, the quantum state is transformed as,
\begin{eqnarray}
\fl |W \rangle^{AB} &= \frac{1}{\sqrt{3}} \bigg(\cos r_1 (|0_{k} \rangle_\mathrm{I}^{+} |0_{-k} \rangle_\mathrm{II}^{-} |1_{k} \rangle_\mathrm{I}^{+} |0_{-k} \rangle_\mathrm{II}^{-} |0\rangle
+ \cos r_2 |0_{k} \rangle_\mathrm{I}^{+}  |0_{-k} \rangle_\mathrm{II}^{-} |0_{k} \rangle_\mathrm{I}^{+}  |0_{-k} \rangle_\mathrm{II}^{-}|1\rangle) \nonumber \\
\fl &+ \cos r_2|1_{k} \rangle_\mathrm{I}^{+}  |0_{-k} \rangle_\mathrm{II}^{-} |0_{k} \rangle_\mathrm{I}^{+}  |0_{-k} \rangle_\mathrm{II}^{-} |0\rangle 
+ e^{-2i \phi} \sin r_1 \sin r_2 |1_{k} \rangle_\mathrm{I}^{+}  |1_{-k} \rangle_\mathrm{II}^{-}|1_{k} \rangle_\mathrm{I}^{+}  
|1_{-k} \rangle_\mathrm{II}^{-} |1 \rangle \nonumber \\
\fl &+ e^{-i\phi} \big( \cos r_1 \sin r_2 |0_{k} \rangle_\mathrm{I}^{+}  |0_{-k} \rangle_\mathrm{II}^{-} |1_{k} \rangle_\mathrm{I}^{+}  |1_{-k} \rangle_\mathrm{II}^{-} |1\rangle +  \cos r_2 \sin r_1 |1_{k} \rangle_\mathrm{I}^+ |1_{-k} \rangle_\mathrm{II}^{-} |0_{k} \rangle_\mathrm{I}^+ |0_{-k} \rangle_\mathrm{II}^- |1 \rangle \nonumber \\
\fl &+ \sin r_1 |1_{k} \rangle_\mathrm{I}^{+} |1_{-k} \rangle_\mathrm{II}^{-} |1_{k} \rangle_\mathrm{I}^{+} |0_{-k} \rangle_\mathrm{II}^{-} 
|0\rangle +  \sin r_2|1_{k} \rangle_\mathrm{I}^{+}  |0_{-k} \rangle_\mathrm{II}^{-} |1_{k} \rangle_\mathrm{I}^{+}  |1_{-k} \rangle_\mathrm{II}^{-} |0\rangle \big ) \bigg)~.
\label{w_ab_accel}
\end{eqnarray}
The eigenvalues of the qubits are:
\begin{eqnarray}
\fl \lambda_{ABC} &= \Bigg\{\frac{\sin^2r_1\sin^2r_2}{3}~, \frac{10+6(\cos2r_1+\cos2r_2)+\cos2(r_1-r_2)+\cos2(r_1+r_2)}{24}~, \nonumber \\
\fl &\frac{6-2(\cos2r_1-\cos2r_2)-\cos2(r_1-r_2)-\cos2(r_1+r_2)}{24} \nonumber \\ 
\fl &\pm \frac{\sqrt{32+8\cos4r_1 -16\cos2r_1\cos^2r_2-16\cos2r_2+8\cos4r_2}}{24}\Bigg\}~;\nonumber \\
\fl \lambda_C &= \left\{\frac{2}{3}, \frac{1}{3}\right\}~; \nonumber \\
\fl \lambda_{AC} &= \left\{\frac{\cos^2r_1}{3}~, \frac{4\pm\sqrt{14+2\cos4r_1}}{12}~,\frac{\sin^2r_1}{3}\right\}~; \nonumber \\
\fl \lambda_{BC} &= \left\{\frac{\cos^2r_2}{3}~, \frac{4\pm\sqrt{14+2\cos4r_2}}{12}~,
\frac{\sin^2r_2}{3}\right\}~. 
\end{eqnarray}
The transformed state can be expressed in terms of the Rindler coordinates in the Minkowski spacetime. Since the two Rindler modes are causally disconnected, we can trace out the modes corresponding to Rindler region-II. We can then compute the total entropy of the system as well as the reduced density matrices. From the expressions for the entropies we find the QCMI to be,
\begin{eqnarray}
\fl I^{AB}_W &= \left(\frac{-\sin^2 r_1 \cos^2 r_2}{3} \right) \log_2 \frac{\sin^2 r_1}{3} - \left(\frac{\cos^2 r_1 \sin^2 r_2}{3} \right) \log_2 \sin^2 r_2 \nonumber \\
\fl &+ \left( \frac{3+\sin^2 r_2}{3}\right) \log_2 3 - \frac{\cos^2 r_1}{3} \log_2 \frac{\cos^2 r_1}{3}
-\frac{\cos^2 r_2}{3} \log_2 \frac{\cos^2 r_2}{3} + \frac{2}{3} \nonumber \\ 
\fl &- \left( \frac{\sqrt{2-\sin^2r_1 \cos^2 r_1 - \sin^2 r_2 \cos^2 r_2}}{3} \right) \log_2 \left( \frac{1+\sqrt{1-\sin^2 r_2 \cos^2 r_2}}{1-\sqrt{1-\sin ^2 r_2 \cos^2 r_2}}\right) \nonumber\\
\fl &- \frac{2}{3} \log_2 \left( \frac{\sin r_1 \sin r_2 \cos r_1 \cos r_2}{9}\right) + \left(\frac{\zeta_2-\sqrt{\zeta_3}}{12}\right) \log_2 \left(\frac{\zeta_2 -\sqrt{\zeta_3}}{12}\right) \nonumber\\
\fl &+ \left(\frac{\zeta_2 + \sqrt{\zeta_3}}{12}\right) \log_2 \left(\frac{\zeta_2 + \sqrt{\zeta_3}}{12} \right)+\frac{\zeta_1}{12} \log_2 \frac {\zeta_1}{12}~,
\end{eqnarray}
with coefficients given by,
\begin{eqnarray}
\fl \zeta_1 &= 5+3 \cos2r_1+3\cos2r_2+\cos2r_1\cos 2r_2~, \nonumber \\
\fl \zeta_2 &= 3-\cos2r_1-\cos2r_2-\cos2r_1\cos2r_2 \nonumber \\
\fl \zeta_3 &= 8-4(\cos2 r_1 + \cos2 r_2) + 2(\cos 4r_1 + \cos 4r_2 -4\cos2r_1\cos2r_2)~.
\end{eqnarray}
In the limit of Bob and Charlie undergoing uniform acceleration, then $r_1 = r_2 = r$. The resultant QCMI is then given by,
\begin{eqnarray}
\fl I^{AB}_W &= \left(\frac{2\sqrt{3+\cos^2 2r}}{3}\right) \log_2 \left(\frac{2-\sqrt{3+\cos^2 2r}}{\sin 2r}\right)
- \left(\frac{3-\sin^2 r \cos^2 r}{3}\right) \log_2 \cos r \sin r\nonumber \\
\fl &+ \left(\frac{\sin^2 r (2+ \cos^2 r)}{3}\right) \log_2 \left(\frac{\sin^2 r (2+\cos^2 r)}{3}\right)
+\left(\frac{2+\cos^2 r}{3}\right) \log_2 3 + \frac{2}{3}\nonumber \\ 
\fl &- \left(\frac{2\cos^2 r +\cos^4 r}{3}\right) \log_2 \left(\frac{2\cos^2 r+\cos^4 r}{3}\right)\nonumber \\ 
\fl &- \left(\frac{4\cos^2 r}{3}\right) \left(\log_2 \cos r + \sin^2 r \log_2 \sin r\right)~.
\label{w_ab_accel_qcmi} 
\end{eqnarray}
From Fig.~(\ref{2qubitaccel}{\color{red}{b}})  we observe a decrease of QCMI with acceleration.  

\section{Analysis}
\label{analysis}

\begin{figure*}[!htb]
\includegraphics[width=\linewidth]{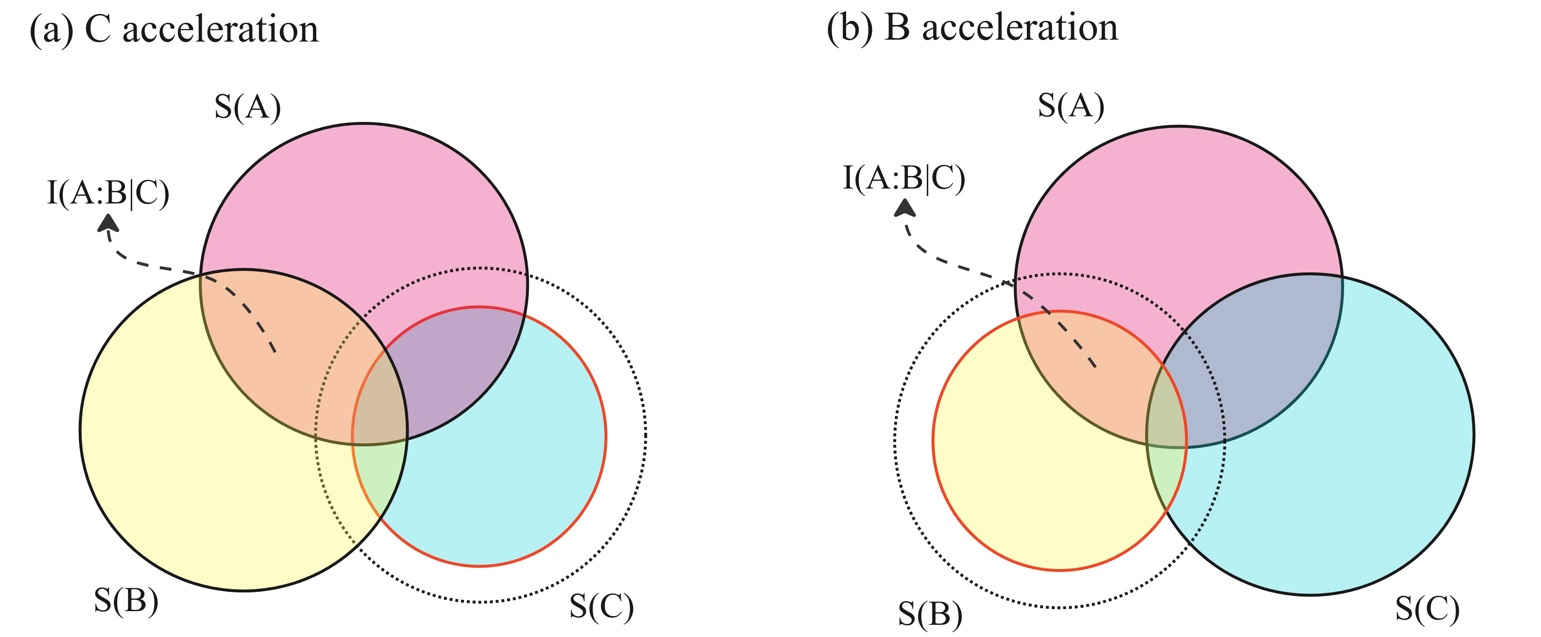}
\caption{The variation of QCMI $(I)$, entropies of two qubit $(S(\rho_{AC}), S(\rho_{BC}))$ and one qubit $(S(\rho_C))$ reduced systems in terms of 
Venn diagram representation is given for $W$ state under (a) C acceleration and (b) B acceleration. Here the dotted circles for qubit C in (a) and for qubit B in (b) represent the qubits before acceleration and the solid circles describe the qubits under acceleration.}
\label{1q_accel_venn}
\end{figure*}

To analyze the qualitative behavior of QCMI we plot the entropies $S(\rho_{AC})$, $S(\rho_{BC})$, $S(\rho_C)$ and $S(\rho_{ABC})$
alongside the QCMI. The QCMI is asymmetric with respect to the roles of Alice, Bob and Charlie. In particular the role of Charlie is to provide 
conditioning information to measure the correlations between Alice and Bob. So in the case of single party acceleration, the cause for the variation in 
QCMI is different depending on whether the accelerating party is Charlie or Bob. When Charlie undergoes acceleration, the amount of information available 
to estimate the correlations between Alice and Bob decreases. We can observe this from Fig. \ref{1q_accel_venn} (a) where we can see the decrease 
in overlap between $C$ and the pair $AB$.
The area (orange shaded) with dotted arrow representing $I(A:B|C)$ gives the correlation of qubits A and B (solid lines) and C (with dotted and solid lines for the unaccelerated and accelerated situations respectively). During Charlie's acceleration his solid line reduces from the original one by the amount of trace of dots. Hence with the new solid line of qubit C it is visible that the area with $I(A:B|C)$ extends in Fig.~\ref{1q_accel_venn} (a). This shows that the correlation between A and B increases and so as the QCMI under Charlie's acceleration.

Likewise the correlations between Alice and Bob also decreases, when Bob undergoes acceleration. The Venn diagrammatic representation of this can be can be seen in Fig. \ref{1q_accel_venn} (b) where we again see a decrease in the overlap between $C$ as well as $AB$.  Here we also see a decrease in the overlap between $A$ and $B$ which corresponds to the decrease in the QCMI as observed in Fig. \ref{1qubitaccel} (b). 
This can be viewed in Fig.~\ref{1q_accel_venn} (b) where the area (orange shaded portion) with dotted arrow denoting $I(A:B|C)$ shrinks due to Bob's acceleration. Hence the correlation between qubits A and B naturally decreases under Bob's acceleration while qubits A and C remains inertial. As a result of this the corresponding QCMI decreases.

\begin{figure*}[!htb]
\includegraphics[width=\linewidth]{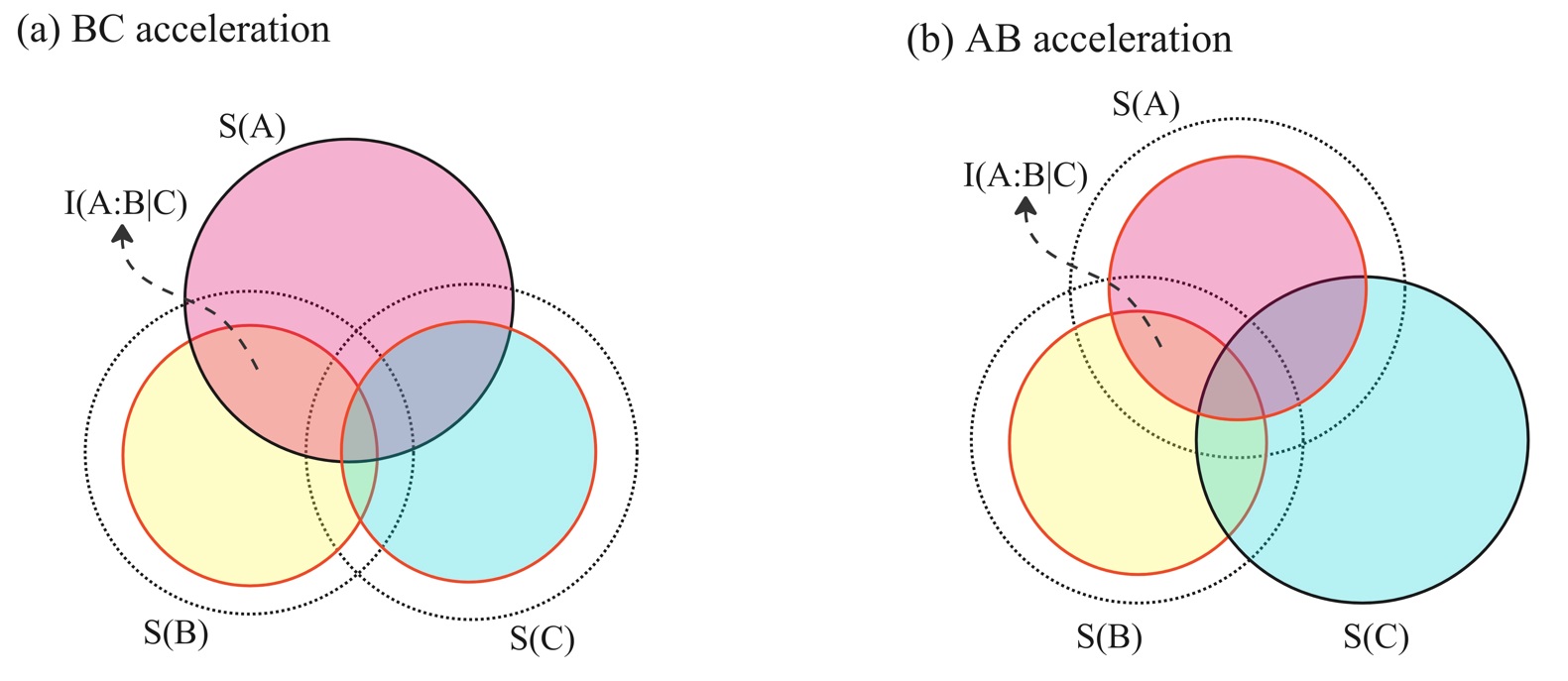}
\caption{The pictorial representation of variation of QCMI $(I)$, entropies of two qubit  $(S(\rho_{AC}), S(\rho_{BC}))$ and one 
qubit $(S(\rho_C))$ reduced systems is given for $W$ state under (a) BC acceleration and (b) AB acceleration.Here the dotted circles for qubits B and C in (a) and for qubits A and B in (b) represent the qubits before acceleration and the solid circles describe the qubits under acceleration.}
\label{2q_accel_venn}
\end{figure*}

Similarly, in the two qubit acceleration scenario, the qualitative change in the entropies is different depending on whether the acceleration is experienced by the pair Bob and Charlie or the pair Alice and Bob. A representation of the two qubit acceleration using Venn diagram is given in Fig. \ref{2q_accel_venn}.  Here Fig. \ref{2q_accel_venn} (a) describes the situation when qubits $B$ and $C$ are accelerated and Fig. \ref{2q_accel_venn} (b) is for the situation when qubits $A$ and $B$ are accelerated. 
From Fig.~\ref{2q_accel_venn} (a) \& (b) it is clear that the QCMI $I(A:B|C)$ increases or decreases depending upon the acceleration of choice of qubits  similar to the observation in single qubit acceleration.
Thus, in total we have four possible situations, 
two each for single qubit and two qubit accelerations. These four possibilities can be divided into two broad categories {\it viz}, 
{\it (a)} when Charlie is being accelerated (i.e qubit $C$ undergoes acceleration and the qubit pair $BC$ undergoes acceleration) and 
{\it (b)} the cases of Bob under acceleration and the pair Alice and Bob under acceleration (qubit $C$ is not under acceleration).  

First we examine the situation {\it (a)} where the qubit $C$ is under 
acceleration and find that the entropies have identical features. The total entropy $S(\rho_{ABC})$ increases monotonically with acceleration and so do the bipartite entropies $S(\rho_{AC})$ and $S(\rho_{BC})$. The entropies increase because the quantum state undergoes decoherence due to relativistic motion. On the contrary, the entropy $S(\rho_C)$ increases to attain a maximum and then decreases. This may be reasoned as follows. When there is no acceleration, the reduced state is $\rho_{c} = \frac{2}{3} |0 \rangle \langle 0 | + \frac{1}{3} |1 \rangle \langle 1|$. Under acceleration this 
state is transformed into $\rho_{c} = \frac{1}{3} (2 \cos^{2} r  |0 \rangle \langle 0 | + (1 + 2 \sin^{2} r ) |1 \rangle \langle 1 |$ where $r$ characterises acceleration. For a $W$-state, the initial amplitude 
$(r=0)$ is $2/3$ for the ground state and $1/3$ for the excited state. Under acceleration, the amplitude gradually shifts more towards the excited state and at $r=\pi/6$ it has a configuration where both the ground and the excited state have equal probability. Hence the entropy increases from $r=0$ (initial state) to $r=\pi/6$ where it is a maximally mixed state. Between $r=\pi/6$ and $r=\pi/4$ the excited state amplitude continues to increase. The state becomes 
$\rho_{c} = \frac{1}{3} |0 \rangle \langle 0 | + \frac{2}{3} |1 \rangle \langle 1|$ at $r=\pi/4$, consequently the entropy decreases. Next we observe the entropies in the case {\it (b)} where the qubit $C$ is not under acceleration. The entropies $S(\rho_{AC})$, $S(\rho_{BC})$ and $S(\rho_{ABC})$ increase with acceleration but $S(\rho_C)$ remains constant. This is because qubit $C$ is not under acceleration.  

When only qubit $C$ is accelerated the QCMI increases. This is because the quantum state reconstruction is harder since there is a loss of conditioning information which was in Charlie's possession. In the remaining cases, i.e, 
{\it (a)} qubit $B$, {\it (b)} qubits $B$ and $C$ are under non-inertial motion and {\it (c)} qubits $A$ and $B$ undergo acceleration, the QCMI decreases. Here the decrease in QCMI does not imply that there is an ease in quantum state reconstruction. It only means that the correlation between Alice and Bob has decreased.

To understand the connection between correlated systems and QCMI we investigate the separable and bi-separable states. A three qubit state is bi-separable when two of its qubits 
are pairwise entangled and the third qubit is separable. There are three possible configurations of bi-separable states, 
$|\phi \rangle_{AB} \otimes |\chi \rangle_{C}$ , $|\phi \rangle_{AC} \otimes |\chi \rangle_{B}$ and $|\phi \rangle_{BC} \otimes |\chi \rangle_{A}$, 
where $|\phi \rangle$ is the maximally entangled Bell state and 
$| \chi \rangle$ is the single qubit state. The explicit forms of the three bi-separable states are,
\begin{eqnarray}
|\varphi_1 \rangle &= \frac{1}{\sqrt{2}} (|0_{A} 0_{B} 0_{C} \rangle  + |1_{A} 1_{B} 0_{C}),  
\label{bisep_AB} \\
|\varphi_2 \rangle &= \frac{1}{\sqrt{2}} (|0_{A} 0_{B} 0_{C} \rangle  + |1_{A} 0_{B} 1_{C}),  
\label{bisep_AC} \\
|\varphi_3 \rangle &= \frac{1}{\sqrt{2}} (|0_{A} 0_{B} 0_{C} \rangle  + |0_{A} 1_{B} 1_{C}).  
\label{bisep_BC}
\end{eqnarray}

Initially all three parties are at rest. When Charlie undergoes acceleration, the quantum states are transformed due to the non-inertial motion. The transformed states are expressed in terms of the Rindler coordinates in the Minkowski spacetime. There are two Rindler modes corresponding to the tripartite quantum states. These modes are causally disconnected and hence we trace out the modes in Rindler region-II. The resultant density matrix is,
\begin{eqnarray}
\fl \rho_{\varphi_1}  &= \frac{1}{2}( \cos^2 r (|000 \rangle \langle 000|+ |000\rangle \langle 110|  + |110 \rangle \langle 110|) + \sin^2 r (|001\rangle \langle 001| \nonumber \\ 
\fl &+  |001 \rangle \langle 111| + |111\rangle \langle 001| + |111 \rangle \langle 111|) + (h.c.)_{od} ),  \\
\fl \rho_{\varphi_2}  &= \frac{1}{2} ( \cos^2 r |000 \rangle \langle 000| + |101 \rangle \langle 101| +\sin^2 r |001\rangle \langle 001|  \nonumber \\
\fl &+ \cos r|000 \rangle \langle 101| + (h.c.)_{od} ), \\ 
\fl \rho_{\varphi_3}  &= \frac{1}{2} ( \cos^2 r |000 \rangle \langle 000| + |011 \rangle \langle 011| +\sin^2 r |001\rangle \langle 001| \nonumber \\  
\fl&+ \cos r|000 \rangle \langle 011| + (h.c.)_{od})~. 
\label{bisep_caccel_dm}
\end{eqnarray}
The total entropy and the reduced state entropies of the bi-separable state Eq.~(\ref{bisep_AB}) are,
\begin{eqnarray}
\fl S(\rho_{ABC}) = S(\rho_C) &= - \sin^2 r \log_2 \sin^2 r 
 - \cos^2 r \log_2 \cos^2 r, \nonumber \\
\fl S(\rho_{AC}) = S (\rho_{BC}) &= 1- \sin^2 r \log_2 \sin^2 r 
 -\cos^2 r \log_2 \cos^2 r~.  
\label{bisep_caccel_entropy}
\end{eqnarray}
The entropies for the density matrices corresponding to the bi-separable states in Eqs.~(\ref{bisep_AC}) \& (\ref{bisep_BC}) are,
\begin{eqnarray}
\fl S(\rho_{ABC}) = S(\rho_{AC}) = - \sin^2 r \log_2 \sin^2 r 
- (1 + \cos^{2}r) \log_{2} (1 + \cos^{2}r),  \nonumber \\
\fl S(\rho_{BC}) = S(\rho_C) = -(1+\sin^2 r) \log_2 (1+\sin^2 r)- \cos^2 r \log_2 \cos^2 r~.  
\label{bisep_ab_ac}
\end{eqnarray}
From the knowledge of the entropy of the total system and the entropies of the subsystems, we find the QCMI to be,
\begin{eqnarray}
 I_{\varphi_1} &= 2;\qquad  I_{\varphi_2} &= I_{\varphi_3}= 0~.
 \label{bisep_caccel_qcmi}
\end{eqnarray}
In the bi-separable quantum state $|\varphi_{1} \rangle$, where qubits $A$ and $B$ are correlated and qubit $C$ is separable, the correlation between $A$ and $B$ is not affected by the accelerated motion of qubit $C$. Since $A$ and $B$ are completely correlated, the QCMI is maximum, remaining so under acceleration. For the bi-separable state $|\varphi_{2} \rangle$ and 
$|\varphi_{3} \rangle$ the qubits $A$ and $B$ are separable. Since the correlation is being measured between the qubits, it will be zero. A completely separable state which can be written as $\rho_{ABC} = \rho_{A} \otimes \rho_{B} \otimes \rho_{C}$ has zero QCMI, since there is no entanglement or correlation between the parties. Thus a decrease of QCMI means that the correlations between qubits $A$ and $B$ are decreasing.  

Qubits under uniform acceleration will experience a loss of quantumness due to decoherence. This loss of information will reflect in the measured values of the QCMI. When qubit $C$ is moving in a non-inertial frame, the QCMI increases because of the conditioning information in it (qubit $C$) decreases making the reconstruction of quantum states harder, since the QCMI represents the difficulty of reconstructing the quantum state.  Here the increase of QCMI does not imply an increase in the quantum correlation between qubits $A$ and $B$, rather it is just the difficulty of reconstruction of the correlated state. In the remaining cases, the QCMI decreases because at least one of either Alice or Bob's qubit is under acceleration. 

\section{Summary and conclusion}
\label{conclusion}

The quantum conditional mutual information (QCMI) of a tripartite $W$-state, when some of the qubits are under 
acceleration is characterised in the present work. A tripartite state for which $I(A:B|C) = 0$ is a quantum Markov 
chain, i.e the tripartite state $\rho_{ABC}$ can be reconstructed using a recovery map $\tau_{C \rightarrow BC}$.
A non-zero finite value of $I(A:B|C)$ provides an upper bound on the distance to the closest reconstructed state. We 
considered the $W$-state under both single qubit acceleration and two qubit acceleration. In QCMI there are two types
of qubits viz {\it(i)} the qubit $C$ which carries the conditioning information and {\it(ii)} the qubit $A$, $B$ whose 
correlations are computed using qubit $C$ (conditioning qubit).

For single qubit acceleration, we consider the two situations 
{\it viz} {\it (a)} qubit $C$ is accelerated and {\it (b)} qubit $B$ is under uniform acceleration. Upon 
acceleration, qubit $C$ loses the information it possess due to decoherence resulting in an increase in QCMI. In the second case, the QCMI decreases 
because the correlation between $A$ and $B$ is decreasing due to acceleration, but the information in $C$ remains constant as observed through the 
value of the entropy $S(\rho_{C})$. Next we investigated the scenario when two of the three qubits undergo acceleration. The two possibilities we 
have are {\it (i)} qubits $B$ \& $C$ under acceleration and {\it (ii)} qubits $A$ \& $B$ are under acceleration. Under both conditions the QCMI 
decreases with acceleration due to decrease in correlation. But the fall in QCMI is faster under the condition {\it (ii)} since both the 
qubits $A$ and $B$ undergo relativistic decoherence. From this result we conclude that the quantum correlations between the qubits held by 
Alice and Bob remain constant when only Charlie is accelerated. But when Bob is accelerated the quantum correlations decrease. In any case there 
are no situations where correlations increase since acceleration manifests itself via decoherence. The QCMI is a very versatile information theoretic 
quantity and its use in the study of the Slepian-Wolf protocol \cite{Devtak2008exactcost} needs to be understood, which will be the basis for our future work. 

\section*{Acknowledgements}
CR was supported in part by a seed grant from IIT Madras to the Centre for Quantum Information, Communication and Computing.

\bibliography{reference}

\providecommand{\newblock}{}
\begin{thebibliography}{10}
\expandafter\ifx\csname url\endcsname\relax
  \def\url#1{{\tt #1}}\fi
\expandafter\ifx\csname urlprefix\endcsname\relax\def\urlprefix{URL }\fi
\providecommand{\eprint}[2][]{\url{#2}}

\bibitem{Unruh1984inertial}
{W~G~Unruh and R~M~Wald} 1984 {\em \em Phys. Rev. D \/\/} {\bf \bf 29} 1047

\bibitem{Dowling2008inertial}
{M~Han, S~J~Olson, and J~P~Dowling} 2008 {\em \em Phys. Rev. A \/\/} {\bf \bf
  78} 022302

\bibitem{Nicolai2010inertial}
{N~Friis, R~A~Bertlmann, M~Huber, and B~C~Hiesmayr} 2010 {\em \em Phys. Rev. A
  \/\/} {\bf \bf 81} 042114

\bibitem{Lanzagobookinertial}
{M~Lanzagorta} 2014 {\em ``\em Quantum Information in Gravitational Fields"\/}
  (Morgan \& Claypool Publishers)

\bibitem{ZhouQKD2018}
{J~Zhou, R~Guo and Y~Guo} 2018 {\em \em Quantum Inf Process\/} {\bf \bf 17} 47

\bibitem{Pierini2018}
{Pierini~R} 2018 {\em Phys. Rev. D\/} {\bf 98}(12) 125007

\bibitem{Hwang2001entanglement}
{M~-R~Hwang, D~Park, and E~Jung} 2001 {\em \em Phys. Rev. A \/\/} {\bf \bf 83}
  012111

\bibitem{Fuentes2005entanglement}
{I~F-Schuller and R~B~Mann} 2005 {\em \em Phys. Rev. Lett. \/\/} {\bf \bf 95}
  120404

\bibitem{Fuentes2006singlemode}
{P~M~Alsing, I~F-Schuller, R~B~Mann and T~E~Tessier} 2006 {\em \em Phys. Rev. A
  \/\/} {\bf \bf 74} 032326

\bibitem{Adesso2007entanglement}
{G~Adesso, I~F-Schuller and M~Ericsson} 2007 {\em \em Phys. Rev. A \/\/} {\bf
  \bf 76} 062112

\bibitem{Dehnavi2011entanglement}
{H~M~Dehnavi, B~Mirza, H~Mohammadzadeh and R~Rahimi} 2011 {\em \em Annals of
  Phys. \/\/} {\bf \bf 326} 1320

\bibitem{Hwang2012entanglement-singlemode}
{M~-R~Hwang, D~Park and E~Jung} 2012 {\em \em Class. Quantum Grav. \/\/} {\bf
  \bf 29} 224004

\bibitem{Ramzan2012entanglement}
{M~Ramzan and M~K~Khan} 2012 {\em \em Quantum Inf Process \/\/} {\bf \bf 11}
  443

\bibitem{Wang2017entanglement}
{W~Y~Sun, D~Wang, J~Yang and L~Ye} 2017 {\em \em Quantum Inf Process \/\/} {\bf
  \bf 16} 90

\bibitem{Jing2018entanglement}
{J~Wang and J~Jing} 2011 {\em \em Phys. Rev. A \/\/} {\bf \bf 83} 022314;
  Erratum 2018 { Phys. Rev. A \/}{\bf 97} 029902

\bibitem{Dong2019wstate}
{A~J~T-Arenas, Q~Dong, G~-H~Sun, W~-C~Qiang and S~- H~Dong} 2019 {\em \em Phys.
  Lett. B \/\/} {\bf \bf 789} 93

\bibitem{Baumgratz2014coherence}
{T~Baumgratz, M~Cramer and M~Plenio} 2014 {\em \em Phys. Rev. Lett. \/\/} {\bf
  \bf 113} 140401

\bibitem{Wu2021coherence}
{S~-M~Wu, H~-S~Zeng and H~-M~Cao} 2021 {\em \em Class. Quantum Grav. \/\/} {\bf
  \bf 38} 185007

\bibitem{Savee2022coherence}
{S~Harikrishnan, S~Jambulingam, P~P~Rohde, and C~Radhakrishnan} 2022 {\em \em
  Phys. Rev. A \/\/} {\bf \bf 105} 052403

\bibitem{Wu2022coh}
{S-M~Wu, H-S~Zeng and H-M~Cao} 2021 {\em \em Class. Quantum Grav. \/\/} {\bf
  \bf 38} 185007

\bibitem{Animesh2009discord}
{A~Datta} 2009 {\em \em Phys. Rev. A \/\/} {\bf 80} 052304

\bibitem{Wang2010discord}
{J~Wang, J~Deng and J~Jing} 2010 {\em \em Phys. Rev. A \/\/} {\bf \bf 81}
  052120

\bibitem{Ramzan2014discordBSMA}
{M~Ramzan} 2014 {\em \em Quantum Inf Process \/\/} {\bf \bf 13} 259

\bibitem{Sugumi2016discord}
{S~Kanno, J~P~Shock, and J~Soda} 2016 {\em \em Phys. Rev. D \/\/} {\bf \bf 94}
  125014

\bibitem{Martinez2010corr}
{E~M-Martínez and J~León} 2010 {\em \em Phys. Rev. A \/\/} {\bf \bf 81}
  032320

\bibitem{Martinez2010boscorr}
{E~M-Martínez and J~León} 2010 {\em \em Phys. Rev. A \/\/} {\bf \bf 81}
  052305

\bibitem{Adesso2012corr}
{G~Adesso, S~Ragy and D~Girolami} 2012 {\em \em Class. Quantum Grav. \/\/} {\bf
  \bf 29} 22

\bibitem{Dragan2013localprojective}
{A~Dragan, J~Doukas, E~M-Martínez and D~E~ Bruschi} 2013 {\em \em Class. and
  Quantum Grav.\/} {\bf \bf 30} 235006

\bibitem{Wang2016quantumsteering}
{W~Jieci, C~Haixin, J~Jiliang and F~Heng} 2016 {\em \em Phys. Rev. D\/} {\bf
  \bf 93} 125011

\bibitem{Zurek2001discord}
{H~Ollivier and W~H~Zurek} 2001 {\em \em Phys. Rev. Lett. \/\/} {\bf \bf 88}
  017901

\bibitem{Vedral2001discord}
{L~Henderson and V~Vedral} 2001 {\em \em J. Phys. A: Mathematical and
  General\/} {\bf \bf 34} 6899

\bibitem{Brandao2015redistribute}
{F~G~S~L~Brandão, A~W~Harrow, J~Oppenheim and S~Strelchuk} 2015 {\em \em Phys.
  Rev. Lett. \/\/} {\bf \bf 115} 050501

\bibitem{Devtak2008exactcost}
{I~Devetak and J~Yard} 2008 {\em \em Phys. Rev. Lett. \/\/} {\bf \bf 100}
  230501

\bibitem{Chandra2020CMI}
{R~Chandrashekar and M~Lauri\`ere, and T~Byrnes} 2020 {\em \em Phys. Rev.
  Lett.\/} {\bf 124} 110401

\bibitem{Brushi2010BSMA}
{D~E~Bruschi, J~Louko, E~M-Martínez, A~Dragan, and I~Fuentes} 2010 {\em \em
  Phys. Rev. A \/\/} {\bf \bf 82} 042332

\bibitem{Soffel1980Bougli}
{M~Soffel, B~Muller and W~Greiner} 1980 {\em \em Phys. Rev. D \/\/} {\bf \bf
  22} 1935

\bibitem{Takagi-1986Bougli}
{S~Takagi} 1986 {\em \em Prog. Theor. Phys. Suppl \/\/} {\bf \bf 88} 1

\bibitem{Jaur1991Bougli}
{R~J\`aureguei, M~Torres and S~Hacyan} 1991 {\em \em Phys. Rev. D \/\/} {\bf
  \bf 43} 3979

\bibitem{Thomas1991Inftheory}
{T~M~Cover and J~A~Thomas} 1991 {\em \em ``Elements of Information Theory"\/}
  (John Wiley \& Sons)

\bibitem{Bennett1999QMI}
{C~H~Bennett, P~W~Shor, J~A~Smolin and A~V~Thapliyal} 1991 {\em \em Phys. Rev.
  Lett. \/\/} {\bf \bf 83} 3081

\bibitem{Jaeger2006Quantuminformation}
{G~Jaeger} 2007 {\em ``\em Quantum Information- An Overview"\/} (Springer, New
  York)

\bibitem{Omar2015qcmiMarkov}
{O~Fawzi and R~Renner } 2015 {\em \em Commun. Math. Phys. \/\/} {\bf \bf 340}
  575

\bibitem{Huang2021nonmarkov}
{Z~Huang and X-K~Guo} 2021 {\em \em Phys. Rev. A \/\/} {\bf \bf 104} 032212

\bibitem{Omar2016recoverymap}
{S~David, O~Fawzi and R~Renner} 2016 {\em \em Proc. R. Soc. A \/\/} {\bf \bf
  472} 20150623

\end{thebibliography}

\begin{appendices}
\section{Method to compute the Quantum Conditional Mutual Information - Single qubit acceleration}
The quantum conditional mutual information for a tripartite state when some of the qubits are accelerated is calculated in the following way:
\begin{itemize}
\item{The tripartite state for which QCMI has to be evaluated is initially at rest.  Here the vacuum and excited states are represented in 
the Minkowski space.}
\item{Under acceleration, the accelerating qubit's vacuum and excited states are to be replaced with the new states represented by the two casually disconnected Rindler regions. 
The vacuum and excited states  corresponding to the single mode approximation are \cite{Fuentes2006singlemode}:
\begin{eqnarray}
|0_k \rangle ^+_M &= \cos r |0_\Omega \rangle^+_\mathrm{I} |0_{-\Omega} \rangle^-_\mathrm{II} + e^{-i\phi}\sin r |1_{\Omega} \rangle^+_\mathrm{I} |1_{-\Omega}\rangle^-_\mathrm{II}, \nonumber \\
|1_k \rangle ^+_M %&= a^\dagger |0_k \rangle^+_M \nn \\ 
&= |1_{\Omega} \rangle^+_\mathrm{I} |0_{-\Omega}\rangle^-_\mathrm{II}~. \nonumber
\label{SMA}
\end{eqnarray}}
\item{The density matrix for the entire system with one among three qubits under acceleration is given by 
\begin{equation}
\rho_W = \ket{W_{ABC}}\bra{W_{ABC}}~. \nonumber \\
\end{equation}
In this tripartite state, there are states corresponding to the Rindler region I as well as Rindler region II.  Since these 
two regions are causally disconnected, we need to trace out one of them.  In our work we trace out Rindler region II
}
\item{From the tripartite density matrix $\rho$ we calculate the reduced density matrices $\rho_{AB}$, $\rho_{BC}$, $\rho_{AC}$ and $\rho_C$ by tracing out the necessary qubits from tripartite state using the formula $\rho_{A} = {\rm Tr}_{BC} \rho_{ABC}$.}
\item{The eigenvalues of the full density matrix as well as the reduced density matrices are computed. Using these eigenvalues 
one can compute the corresponding entropies using the entropic relation:
\begin{equation}
 S(\rho) = - \sum_i \lambda_i \log\lambda_i~, \nonumber \\
\end{equation}
where $\lambda_i$'s are the eigenvalues of density matrix under consideration.}
\item{From the entropies, the QCMI is calculated using the relation:
\begin{equation}
I(A:B|C) = S(\rho_{AC}) + S(\rho_{BC}) - S(\rho_{ABC})-S(\rho_{C})~.\nonumber \\
\end{equation}
The QCMI measures the correlations between $A$ and $B$ from the point of view of $C$. 
}
\end{itemize}
\end{appendices}

\end{document}